\documentclass[%
 aip,
 amsmath,amssymb,
 reprint,%
]{revtex4-2}

\usepackage{graphicx}
\usepackage{dcolumn}
\usepackage{bm}
\usepackage{amsmath}
\usepackage{braket}

\usepackage{comment}
\usepackage[utf8]{inputenc}
\usepackage[T1]{fontenc}
\usepackage{mathptmx}
\usepackage{etoolbox}
\usepackage{xcolor}

\newcommand{\vect}[1]{\mathbf{#1}}

\newcommand{\bmat}[1]{\mathbf{#1}}

\newcommand{\kp}[1]{\mathbf{#1}}
\newcommand{\spin}[0]{ }

\definecolor{ykred}{rgb}{1.0, 0.0, 0.0}
\newcommand{\yk}[1]{{\color{ykred}{#1}}}
\definecolor{jxblue}{rgb}{0.0, 0.0, 1.0}
\newcommand{\jx}[1]{{\color{jxblue}{#1}}}
\definecolor{csviolet}{rgb}{1.0,0.0,1.0}

\makeatletter
\def\@email#1#2{%
 \endgroup
 \patchcmd{\titleblock@produce}
  {\frontmatter@RRAPformat}
  {\frontmatter@RRAPformat{\produce@RRAP{*#1\href{mailto:#2}{#2}}}\frontmatter@RRAPformat}
  {}{}
}%
\makeatother

\begin{document}


\title[]
{All-electron RT-TDDFT with Exact Exchange based on Numeric Atom-centered Orbitals for Condensed Matter}

\author{Jianhang Xu}
\altaffiliation{These authors contributed equally to this work.}
\affiliation{Department of Chemistry, University of North Carolina at Chapel Hill, Chapel Hill, North Carolina 27599, USA}
\author{Christopher Shepard}
\altaffiliation{These authors contributed equally to this work.}
\affiliation{Department of Chemistry, University of North Carolina at Chapel Hill, Chapel Hill, North Carolina 27599, USA}
\author{Yosuke Kanai}
\email{ykanai@unc.edu}
\affiliation{Department of Chemistry, University of North Carolina at Chapel Hill, Chapel Hill, North Carolina 27599, USA}
\affiliation{Department of Physics and Astronomy, University of North Carolina at Chapel Hill, Chapel Hill, North Carolina 27599, USA}

\date{\today}

\begin{abstract}

Real-time time-dependent density functional theory (RT-TDDFT) has become a widely used first-principles method for studying excited-state and nonequilibrium electron dynamics. 
Hybrid exchange-correlation (XC) functionals, which incorporate a fraction of nonlocal exact exchange, are particularly important for accurately describing processes such as charge transfer and hot-carrier dynamics. 
However, the high computational cost which accompanies exact exchange, as required in hybrid XC functionals, has largely restricted RT-TDDFT simulations of condensed matter systems to using semi-local XC functionals. 
In this work, we present an all-electron implementation of exact exchange for RT-TDDFT based on numeric atom-centered orbitals, extending an efficient ground-state resolution-of-identity (RI) framework to the complex, time-dependent Hamiltonians required for real-time propagation. 
We validate the implementation against an established RI approach, for an isolated water molecule, finding exact agreement for both valence and core (K-edge) excitations.
We also demonstrate the method on two extended periodic systems, crystalline silicon and silicon carbide. 
For silicon, the HSE06 spectrum reproduces previously reported real-time TDDFT results and agrees with experiment to within approximately 0.2~eV, consistent with plane-wave pseudopotential implementations. 
For silicon carbide, we compute valence and X-ray absorption spectra at the silicon $L$-edge and carbon $K$-edge. 
These results establish an efficient, all-electron route to hybrid-XC functional RT-TDDFT for both molecular and condensed matter systems.

\end{abstract}

\maketitle

\section{\label{sec:intro} Introduction}

\par 
Understanding and predicting excited-state and nonequilibrium dynamical phenomena is central to modern condensed matter physics, materials science, and chemistry. 
Many of today's challenges in science and technology are directly related to electron dynamics under external stimuli. 
The nonequilibrium response of electrons lies at the heart of a wide variety of key phenomena that may be harnessed to develop new technologies\cite{OGINNI_2026_review, jamshaid2026electronic, Yang_2022_electro}.
Together with advances in ultrafast spectroscopy, first-principles computational methods have made it possible to probe and predict electronic, spin, lattice, excitonic, and other quasiparticle dynamics in materials on unprecedented temporal and spatial scales. 
These recent developments are opening new opportunities to uncover emergent physical phenomena and to design new materials with tailored functionalities.

Over the past fifty years, various first-principles methodologies have been developed to study nonequilibrium electron dynamics\cite{blum2024roadmap, Casida_1995_LR, Ullrich_2025_TDDFT, maitra2016perspective}.
In particular, the explicit real-time propagation approach to time-dependent density functional theory (RT-TDDFT)\cite{Li_2020_RT-TDDFT, hekele2021all, Yabana_1996_RT, Shepard_2021_JCP_RT, Isborn_2016_RT, Xu_2024_RTJacs, kononov2022electron} has emerged as a powerful and computationally efficient framework for investigating ultrafast electron dynamics, especially for studying large condensed matter systems \cite{schleife2014quantum, shepard2023electronic, shepard2023ion, xu2023velocity}. 
RT-TDDFT has become a versatile computational method for studying a broad range of excited-state and nonequilibrium phenomena, including optical absorption,\cite{fernando2015x,yost2019propagation, liu2022algorithm} electronic stopping,\cite{pruneda2007electronic,schleife2015accurate,Reeves2017, yao2019k,yost2019electronic,shepard2023electronic} electronic circular dichroism spectra,\cite{goings2016atomic,mattiat2022comparison} pump-probe spectroscopy, \cite{Qiao_2026_first, Torsha_2023_pump, raya2026exciting} topological quantum matter, \cite{hubener2017creating, zhou2023molecular}, laser-induced water splitting \cite{Yoshiyuki_2017_watersplit} and high-harmonic generation.\cite{tancogne2017impact,mrudul2020high}
The increasing popularity of RT-TDDFT has led to its numerical implementation in a wide variety of electronic structure codes, including SALMON, \cite{noda2019salmon} SIESTA, \cite{soler2002siesta, takimoto2007real} GAUSSIAN, \cite{liang2011efficient, Li_2005_Gaussian} CP2K, \cite{kuhne2020cp2k} Q-Chem, \cite{shao2015advances, zhu2018self} INQ, \cite{andrade2021inq} QB@LL/QB@CH, \cite{schleife2012plane, Shepard_2024_EXX} NWChem, \cite{lopata2011modeling, valiev2010nwchem} Octopus, \cite{tancogne2020octopus, marques2003octopus} TDAP, \cite{lian2018photoexcitation, ren2010optical} ELK, \cite{elk_code}, Exciting \cite{Rodrigues_2021_exciting}, 
\textsc{FHI-aims} \cite{AIMS2026Roadmap}, PySCF\cite{hanasaki2023implementation}, and more. 
Implementation details vary widely between codes, particularly in choice of basis representations, including real-space grids, plane waves, Gaussian-type orbitals, and numerical atomic-centered orbitals (NAOs).
Among these approaches, the \textsc{FHI-aims} electronic structure package \cite{blum2009ab,abbott2025roadmap} provides a highly accurate all-electron NAO framework for RT-TDDFT simulations in both finite and periodic systems\cite{hekele2021all}.

\par
While RT-TDDFT capabilities have continued to expand, the availability of advanced exchange-correlation (XC) functionals has progressed more slowly. Unlike ground-state DFT, where XC approximations with exact exchange and even the random phase approximation have become commonplace \cite{Chen_2017_RPA, Trushin_2025_RPA}, most RT-TDDFT simulations still primarily rely on local and semi-local XC functionals for simulating condensed matter systems. 
These density functional approximations, such as the local density approximation (LDA) \cite{Hohenberg_1964_LDA}, generalized gradient approximations (GGA) \cite{perdew1996generalized, kurth1999molecular}, and meta-GGAs \cite{Kummel_2019_mgga}, suffer from self-interaction errors and an inaccurate description of long-range response. 
As a result, they often provide limited accuracy for systems involving charge localization or charge transfer.\cite{cohen2008insights}
Hybrid XC functionals provide an effective and practical route to mitigate these deficiencies and improve the description of excited-state and time-dependent phenomena.\cite{paier2008dielectric, maitra2017charge, refaely2015solid}
Their key component is the incorporation of a fraction of nonlocal exact (Hartree–Fock) exchange combined with the local or semi-local exchange from an LDA, GGA, or meta-GGA functional.
Although various hybrid XC functional forms with different mixing schemes have been proposed, \cite{janesko2009screened,toulouse2022review} the computational cost of exact exchange (EXX) is often prohibitive (if not insurmountable), particularly for  condensed matter systems. 
The \textsc{FHI-aims} code offers an efficient and scalable framework for evaluating EXX, leveraging the localized NAO basis together with the rich development history of resolution-of-identity (RI) techniques. \cite{whitten1973coulombic, dunlap2000robust, weigend1998ri, vahtras1993integral, kokott2024efficient, kokott2025efficient}
\par In this work, we discuss the implementation of RT-TDDFT with EXX within \textsc{FHI-aims} code, the architecture of the existing RI techniques and the extension of the localized resolution of identity (RI-LVL) method for compatibility with RT-TDDFT. 
To validate our implementation, we compute the optical absorption spectra for an isolated water molecule and demonstrate that both the valence and $K$-edge excitations are reproduced exactly when compared to the standard RI method. 
Finally, we apply our RI-LVL-based hybrid-XC implementation to two extended periodic systems, crystalline silicon and 2H-silicon carbide, highlighting the importance of using hybrid XC functionals in RT-TDDFT simulations.

\section{\label{sec:imp} Theory and computational method}

\subsection{\label{subsec:RT-TDDFT} Real-time time-dependent density functional theory}
The Runge–Gross theorem\cite{RungeGross1984} 
allows us to formulate the Kohn--Sham (KS) ansatz for quantum dynamics such that the exact time-dependent density $\rho(\vect{r},t)$ is given by a set of single-particle orbitals
$\{\psi\spin_{n\kp{k}}(\vect{r},t)\}$,
where $n$ is the band index and $\kp{k}$ is a Bloch wave vector in the first Brillouin zone. 
Their evolution, thus also the temporal change of the density, is given by the time-dependent Kohn--Sham (TDKS) equation
\begin{equation}
\begin{aligned}
  i\frac{\partial}{\partial t}\psi\spin_{n\kp{k}}(\vect{r},t)
   = \left[
     -\frac{1}{2}\right.& \nabla^2 
     + v_{\mathrm{H}}[\rho(\vect{r},t)] \\
     & \left. \vphantom{\frac{1}{2}\nabla^2}
     + v_{\mathrm{XC}}\spin[\rho(\vect{r},t)]
     + v_{\mathrm{ext}}(\vect{r},t)
     \right]\psi\spin_{n\kp{k}}(\vect{r},t)
  \label{eq:TDKS_1}
\end{aligned}
\end{equation}
where  
$v_{\mathrm{H}}$, $v_{\mathrm{XC}}\spin$, and $v_{\mathrm{ext}}$ are the Hartree, exchange-correlation (XC), and external potentials, respectively.  
The spin degree of freedom is omitted throughout this work for brevity.
The TDKS equation (Eq.~\ref{eq:TDKS_1}) is formally exact if the exact XC functional $v_{\mathrm{XC}}\spin$ is known. 
In practical applications, approximate XC functionals are employed together with the \emph{adiabatic approximation}\cite{gross1996m}, in which the XC potential at time $t$ depends only on the instantaneous density
\begin{equation}
  \rho\spin(\vect{r},t)
  = \sum_{n}\int_{\mathrm{BZ}} d\kp{k}\, 
  f\spin_{n\kp{k}}
  \left|\psi\spin_{n\kp{k}}(\vect{r},t)\right|^2,
\end{equation}
where $f\spin_{n\kp{k}}$ denotes the occupation number of the $(n,\kp{k})$ state. 
This approximation neglects memory effects 
but it has proven accurate for a broad range of excitations and dynamical responses.\cite{Isborn_2016_RT}

\par 
One of the most widely-used applications of RT-TDDFT simulation is to study the electronic excitation response to external electromagnetic fields.
The external potential due to the electromagnetic field is included as $v_{\mathrm{ext}}$ in the KS Hamiltonian, and it is often treated within the dipole (long-wavelength) approximation (i.e., spatial variations of the field are neglected and only its time dependence is retained). \cite{bandrauk2013atoms} 
The external potential is given by
\begin{equation}
v_{\mathrm{ext}}(t) = -\, \vect{\hat{r}} \cdot \vect{E}(t),
\label{eq:length_gauge}
\end{equation}
where $\vect{E}(t)$ is the time-dependent electric field and $\vect{\hat{r}}$ is the position operator.
Instead of using this so-called length gauge, it is often convenient to exploit gauge invariance and use another gauge when extended systems are studied because defining the position operator 
with periodic boundary conditions is not straightforward.
In the alternative velocity gauge, the external potential operator reads
\begin{equation}
v_{\mathrm{ext}}(t) = -i\, \vect{A}(t) \cdot \nabla + \frac{1}{2}\vect{A}(t)^2,
\end{equation}
where $\vect{A}(t)$ is the vector potential such that $\vect{E}(t) = -\partial_t \vect{A}(t)$.
The time evolution of the TDKS equation (Eq.~\ref{eq:TDKS_1}) is typically initiated from the ground-state KS orbitals at $t=0$ (this also alleviates the necessity to consider the XC potential dependence on the initial many-body state),\cite{marques2004time,maitra2016perspective} typically under an external perturbation representing a laser (i.e. single frequency) pulse or a $\delta$-kick electric field.
The time-dependent density is propagated to update the Hartree and XC potentials, $v_{\mathrm{H}}$ and $v_{\mathrm{XC}}$, at each time step.

\subsection{\label{subsec:NAO} Formulation with numeric atom-centered orbitals}

\par In \textsc{FHI-aims} code,\cite{blum2009ab,hekele2021all,abbott2025roadmap} the KS orbitals are expanded in a Bloch-adapted basis as
\begin{equation}
    \psi\spin_{n\kp{k}}(\vect{r},t) =
    \sum_{i}
    c\spin_{i n \kp{k}}(t)
    \sum_{\vect{R}}
    e^{i\kp{k}\cdot \vect{R}}
    \phi^{\mathrm{NAO}}_{i(\vect{R})}(\vect{r}),
\end{equation}
where the localized numeric atom-centered orbital (NAO) basis functions at the Bravais lattice vectors $\vect{R}$ are defined as  $\phi^{\mathrm{NAO}}_{i(\vect{R})}(\vect{r}) =
    \phi^{\mathrm{NAO}}_{i}\left(\vect{r} - \vect{R}\right)$.

\par The KS orbitals are propagated in time according to the TDKS equation. 
In the matrix representation, the corresponding equation of motion (EOM) reads
\begin{equation}
\label{eq:eom_p_rttddft}
    \frac{d}{dt}\,\bmat{C}\spin(\kp{k},t) = 
    -i\,\bmat{S}^{-1}(\kp{k})\bmat{H}\spin(\kp{k},t)\bmat{C}\spin(\kp{k},t),
\end{equation}
where $\bmat{C}\spin(\kp{k},t)$ are the expansion coefficients $c\spin_{i n \kp{k}}(t)$, which contain the time-dependence of the electronic system. 
The matrices $\bmat{S}(\kp{k})$ and $\bmat{H}\spin(\kp{k},t)$ are the overlap and Hamiltonian matrices, respectively, in the Bloch-summed NAO basis, defined as
$
S_{ij}(\kp{k}) = \langle \phi_{i\kp{k}} | \phi_{j\kp{k}} \rangle$, and 
$H\spin_{ij}(\kp{k},t) = \langle \phi_{i\kp{k}} | \hat H^{\mathrm{KS}}(t) | \phi_{j\kp{k}} \rangle.
$ 
In practice, \textsc{FHI-aims} constructs $\bmat{S}(\kp{k})$ and $\bmat{H}(\kp{k},t)$ from their real-space representations, $\bmat{S}(\vect{R})$ and $\bmat{H}(\vect{R},t)$,
$
S_{ij}(\vect{R}) = \langle \phi_{i(\vect{0})} | \phi_{j(\vect{R})} \rangle$, and 
$H\spin_{ij}(\vect{R},t) = \langle \phi_{i(\vect{0})} | \hat H^{\mathrm{KS}}(t) | \phi_{j(\vect{R})} \rangle,
$ in the NAO basis:
\begin{equation}
\begin{aligned}
\bmat{S}(\kp{k}) &= \sum_{\vect{R}} e^{i\kp{k}\cdot\vect{R}}\bmat{S}(\vect{R}), \\
\bmat{H}(\kp{k},t) &= \sum_{\vect{R}} e^{i\kp{k}\cdot\vect{R}}\bmat{H}(\vect{R},t). 
\end{aligned}
\end{equation}
The equation of motion (Eq.~\ref{eq:eom_p_rttddft}) can then be solved using a variety of time-propagation schemes.\cite{Castro04propagators, Castro18propagators, Varga17propagators} 
Here, we show an exponential integration scheme,\cite{magnus1954exponential} in which the KS orbitals are propagated over a time step, $\Delta t$, as
\begin{equation}
    \bmat{C}(\kp{k},t+\Delta t)=\exp(\hat O)\,\bmat{C}(\kp{k},t),
\end{equation}
where $\hat O$ denotes the corresponding effective time-evolution operator.
Using the exponential-midpoint (EM) propagator,\cite{Castro18propagators} the time-dependent expansion coefficients are evolved as
\begin{equation}
\begin{aligned}
    \label{eq:eom_p_rttddft_EM}
    \bmat{C}(\kp{k},t+&\Delta t) = 
     \bmat{S}(\kp{k})^{-\frac{1}{2}} \\
    &\exp\!\left[
    -i\Delta t\,
    \bmat{S}(\kp{k})^{-\frac{1}{2}}
    \bmat{H}\!\left(\kp{k},t+\Delta t/2\right)
    \bmat{S}(\kp{k})^{-\frac{1}{2}}
    \right]
    \bmat{S}(\kp{k})^{\frac{1}{2}}
    \bmat{C}(\kp{k},t).
\end{aligned}
\end{equation}
Because the TDKS equations are nonlinear, i.e., the Hamiltonian depends on the electron density and, for hybrid functionals, on the instantaneous nonlocal exchange, the Hamiltonian matrix at time $t+\Delta t/2$ is obtained using an iterative predictor--corrector procedure.\cite{zhu2018self}
The matrix exponential in Eq.~\eqref{eq:eom_p_rttddft_EM} is evaluated using the eigenvector decomposition method,\cite{moler1978nineteen}
\begin{equation}
    \exp(\bmat{A})=
    \bmat{V}\,
    \mathrm{diag}\!\left(e^{\lambda_1},\ldots,e^{\lambda_n}\right)
    \bmat{V}^{-1},
\end{equation}
where $\bmat{V}$ and $\lambda_i$ are the eigenvectors and eigenvalues of the matrix $\bmat{A}$, respectively.
Alternative propagators, including the Crank--Nicolson (CN) and enforced time-reversal symmetry (ETRS) schemes, are also implemented for solving Eq.~\eqref{eq:eom_p_rttddft} in \textsc{FHI-aims}\cite{hekele2021all}.

\subsection{\label{subsec:hybrid} Hybrid Exchange-Correlation functionals}


\par With the most widely adapted global hybrid XC form, the (generalized) TDKS equation becomes
\begin{equation}
\begin{aligned}
  i\frac{\partial}{\partial t}&\psi\spin_{n\kp{k}}(\vect{r},t)
   = \left[
     -\frac{1}{2}\nabla^2
     + v_{\mathrm{H}}[\rho(\vect{r},t)]
     + v\spin_{\mathrm{C}}[\rho(\vect{r},t)] \right.\\
     & \left.
     + (1-\alpha)v\spin_{\mathrm{X}}[\rho(\vect{r},t)\spin]
     - \alpha \hat K\spin(t)
     + v_{\mathrm{ext}}(t)
     \right]\psi\spin_{n\kp{k}}(\vect{r},t),
  \label{eq:TDKS_2}
\end{aligned}
\end{equation}
where $v_{\mathrm{X}}$ and $v_{\mathrm{C}}$ denote the local exchange and correlation potentials, respectively, and $\alpha$ is the mixing parameter.
The exact (Fock) exchange operator $\hat K\spin(t)$ can be expressed as
\begin{equation}
\begin{aligned}
\hat K\spin(t)\,
&\psi\spin_{n\kp{k}}(\vect{r},t)
= \\
&\sum_{m,\kp{q}} f\spin_{m\kp{q}}\,
\psi\spin_{m\kp{q}}(\vect{r},t)
\int \,
\frac{
\psi\spin^{*}_{m\kp{q}}(\vect{r}',t)\,
\psi\spin_{n\kp{k}}(\vect{r}',t)
}{
|\vect{r}-\vect{r}'|
}d\vect{r}'.
\end{aligned}
\end{equation}
Here, both $\vect k$ and $\vect q$ are Bloch wave vectors sampling the first Brillouin zone.
As discussed in Sec.~\ref{subsec:RT-TDDFT}, \textsc{FHI-aims} employs a NAO basis to represent the KS orbitals.
The matrix elements of the exact exchange operator in the NAO Bloch basis are given by
\begin{equation}
\label{eq:exchange_nao_density}
\begin{aligned}
K\spin_{ij}(\kp{k},t)
& \equiv
\braket{\phi_{i\kp{k}}|\hat K\spin(t)|\phi_{j\kp{k}}} \\
& = 
\frac{1}{N_{\kp{q}}}
\sum_{lm,\kp{q}}
D\spin_{lm}(\kp{q},t)
\,
( i l | m j)_{\kp{k},\kp{q}}
\end{aligned}
\end{equation}
where $N_{\kp{q}}$ is the number of $\kp{k}$-points sampled in the first Brillouin zone, and the four-center two-electron Coulomb ($2e$) integrals are 
\begin{equation}
( i l | m j )_{\kp{k},\kp{q}}
=
\iint 
\frac{
\phi^*_{i\kp{k}}(\vect r)
\phi_{l\kp{q}}(\vect r)
\phi^*_{m\kp{q}}(\vect r')
\phi_{j\kp{k}}(\vect r')
}{
|\vect r - \vect r'|
}d\vect r\,d\vect r'
    \label{eq:2e-coulomb}
\end{equation}
and $D_{lm}$ is the $\kp{k/q}$-dependent density matrix 
\begin{equation}
    D_{lm}\spin(\kp{q},t)=\sum_n
    f_{n\kp{q}}
    c_{ln,\kp{q}}\spin(t)
    c_{mn,\kp{q}}\spin^{*}(t).
\end{equation}
The key limitation of using hybrid XC functionals in RT-TDDFT (and DFT) is the substantial computational cost associated with evaluating the nonlocal Fock exchange operator $\hat K\spin(t)$ at every time step. 
In this work, we present a scalable hybrid XC-functional RT-TDDFT implementation in \textsc{FHI-aims}, based on accurate all-electron numeric atom-centered orbitals and an efficient resolution-of-identity (RI) treatment\cite{ihrig2015accurate} of the exchange operator.

In the following, we briefly outline the numerical techniques that underlying the hybrid-functional RT-TDDFT implementation. 
Much of the methodology builds upon previously established hybrid XC-functional developments in \textsc{FHI-aims}.\cite{levchenko2015hybrid,kokott2024efficient,bussy2024efficient}

\subsubsection{\label{subsec:RI} Exchange operator using resolution of identity (RI)}

One major computational challenge in evaluating the exchange operator when using NAO basis functions is representing the $2e$ integral, $( i l | m j )_{\kp{k},\kp{q}}$, efficiently. 
One can express the product of the two basis functions with the same spatial coordinate using a linear combination of an auxiliary basis set.
This approach is often referred to as "resolution of identity (RI)" or "density fitting" in the literature\cite{whitten1973coulombic, Dunlap_1979_RI, dunlap2000robust, vahtras1993integral, weigend1998ri, feyereisen_1993_use}.
In general, with RI, the basis pair is expanded in terms of the auxiliary basis set as
\begin{equation}
    \phi_{i\kp{k}}^*(\vect{r})\phi_{j\kp{q}}(\vect{r})=
    \sum_\mu C_{ij}^\mu(\kp{k},\kp{q}) P_{\mu\kp{(q-k)}}(\vect{r}),
    \label{eq:co_ri}
\end{equation}
where $C_{ij}^\mu(\kp{k},\kp{q})$ are the expansion coefficients.
Here, the auxiliary basis set also carries the Bloch summation,
\begin{equation}
    P_{\mu\kp{k}}(\vect{r}) = 
    \sum_{\vect{R}}
    e^{i\kp{k}\cdot \vect{R}}
    P_{\mu,\vect{N}}(\vect{r}-\vect{R}).
\end{equation}
As a result, the two electron integral can be written as 
\begin{equation}
    ( i l | m j )_{\kp{k},\kp{q}} = 
    \sum_{\mu\nu}
    C_{il}^\mu(\kp{k},\kp{q})
    V_{\mu\nu}(\kp{k}-\kp{q})
    C_{mj}^\nu(\kp{q},\kp{k}),
\end{equation}
where the Coulomb matrix in the auxiliary basis is defined as
\begin{equation}
    V_{\mu\nu}(\kp{k}) = 
    \iint 
    \frac{P^*_{\mu\kp{k}}(\vect{r})
    P_{\nu\kp{k}}(\vect{r'})}
    {\left|\vect{r}-\vect{r}'\right|
    }d\vect{r}\,d\vect{r}'.
\end{equation}
In extended periodic systems, the coulomb matrix $V_{\mu\nu}(\kp{k})$ diverges at $\kp{k}=0$.\cite{gygi1986self,lin2026coulomb}
This singularity can be treated by applying a cutoff for the Coulomb operator beyond a certain distance (the cut-Coulomb operator $\tilde V_{\mu\nu}(\kp{k})$).\cite{spencer2008efficient, levchenko2015hybrid}

The idea of finding the expansion coefficients with RI is to minimize the error due to the residual function 
\begin{equation}
    \delta\rho_{i\kp{k},j\kp{q}}(\vect{r}) = 
    \phi_{i\kp{k}}^*(\vect{r})\phi_{j\kp{q}}(\vect{r}) -
    \sum_\mu C_{ij}^\mu(\kp{k},\kp{q}) P_{\mu(\kp{q}-\kp{k})}(\vect{r}).
\end{equation}
One commonly used strategy is to reduce Coulomb repulsion between the residuals
\begin{equation}
    \left( \delta\rho_{i\kp{k},j\kp{q}}
    | \delta\rho_{i\kp{k},j\kp{q}}\right) =
    \iint 
    \frac{\delta\rho_{i\kp{k},j\kp{q}}^*(\vect{r})
    \delta\rho_{i\kp{k},j\kp{q}}(\vect{r}')}
    {\left|\vect{r}-\vect{r}'\right| }
    d\vect{r}\,d\vect{r}'.
\end{equation}
This is a particularly suitable choice for evaluating the Coulomb integral in Eq. \ref{eq:2e-coulomb}.
The corresponding expansion coefficients are 
\begin{equation}
    C_{ij}^\mu(\kp{k},\kp{q}) = 
    \sum_\nu (ij|\nu)_{\kp{k},\kp{q}}V_{\nu\mu}^{-1}(\kp{q}-\kp{k}),
\end{equation}
where 
\begin{equation}
    (ij|\nu)_{\kp{k},\kp{q}} = \iint 
    \frac{\phi_{i\kp{k}}^*(\vect{r})
    \phi_{j\kp{q}}(\vect{r}) 
     P_{\nu(\kp{q}-\kp{k})}(\vect{r}')}
     {\left|\vect{r}-\vect{r}'\right| }
    d\vect{r}\,d\vect{r}'.
\end{equation}
Given its format, this flavor of RI method is often referred to as the RI-V method.
The nonlocal Fock-exchange matrix in reciprocal space is generally expressed using the RI approach as
\begin{equation}
    \begin{aligned}
    K_{ij}&\spin(\kp{k},t)= \\
    &\frac{1}{N_{\kp{q}}}\sum_{lm,\kp{q}}
    D_{lm}\spin(\kp{q},t)
    \sum_{\mu\nu}
    C_{il}^\mu(\kp{k},\kp{q})
    V_{\mu\nu}(\kp{k}-\kp{q})
    C_{mj}^\nu(\kp{q},\kp{k}).
    \end{aligned}
\end{equation}

\subsubsection{\label{subsec:RILVL-k} Localized RI (RI-LVL)}

For extended periodic systems, the RI method in Eq.~\ref{eq:co_ri} exhibits unfavorable scaling with respect to the number of k-points due to its double dependence on the Bloch wave vector.
To overcome this limitation, \citeauthor{ihrig2015accurate} formulated a localized version of the RI-V method by restricting the auxiliary expansion to a pair domain in real space, which is referred to as RI-LVL.
The RI-LVL implies 
\begin{equation}
\begin{aligned}
    \phi_{i(\vect{R})}&(\vect{r})
    \phi_{j(\vect{R}')}(\vect{r}) = \\
    &\sum_\mu\left(
    C_{i(\vect{R})j(\vect{R}')}^{\mu(\vect{R})}
    P_{\mu(\vect{R})}(\vect{r}) +
    C_{i(\vect{R})j(\vect{R}')}^{\mu(\vect{R}')}
    P_{\mu(\vect{R}')}(\vect{r})
    \right),
\end{aligned}
\label{eq:ap_rilvl}
\end{equation}
where $\vect{R}$ and $\vect{R}'$ are the Bravais lattice vectors, and $i(\vect{R})$ explicitly states that the basis $i$ belongs to cell $\vect{R}$.
Only auxiliary functions that belong to the same cell as the atomic basis functions are used in the expansion.
As in the isolated molecular case, the auxiliary basis function $\mu$ is restricted to be centered on the atom of either basis $i(\vect{R})$ or, $j(\vect{R}')$.
In reciprocal space, we have
\begin{equation}
\begin{aligned}
    \phi_{i\kp{k}}^*&(\vect{r})
    \phi_{j\kp{q}}(\vect{r}) = 
    \sum_{\vect{R},\vect{R}'}
    e^{-i\kp{k}\cdot\vect{R}}
    e^{i\kp{q}\cdot\vect{R}'}
    \phi_{i(\vect{R})}(\vect{r})
    \phi_{j(\vect{R}')}(\vect{r}) \\
    = &  \sum_{\vect{R},\vect{R}'}
    e^{-i\kp{k}\cdot\vect{R}}
    e^{i\kp{q}\cdot\vect{R}'}
    \sum_\mu\left(
    C_{i(\vect{R})j(\vect{R}')}^{\mu(\vect{R})}
    P_{\mu(\vect{R})}(\vect{r}) +
    C_{i(\vect{R})j(\vect{R}')}^{\mu(\vect{R}')}
    P_{\mu(\vect{R}')}(\vect{r})
    \right) \\
    = & \sum_{\vect{R},\vect{R}'}
    e^{-i\kp{k}\cdot\vect{R}}
    e^{i\kp{q}\cdot\vect{R}'}
    \sum_\mu\left(
    C_{i(\vect{0})j(\vect{R}'-\vect{R})}^{\mu(\vect{0})}
    P_{\mu(\vect{R})}(\vect{r}) +
    C_{i(\vect{R}-\vect{R}')j(\vect{0})}^{\mu(\vect{0})}
    P_{\mu(\vect{R}')}(\vect{r})
    \right) \\
    = &\sum_\mu 
    \left( C_{ji}^\mu(\kp{q},\vect{0}) +
    C_{ij}^\mu(\vect{-k},\vect{0})
    \right)
    P_{\mu(\kp{q}-\kp{k})}(\vect{r}),
    \label{eq:co_rilvl}
\end{aligned}
\end{equation}
where $\vect{0}$ denotes the origin unit cell, and the reciprocal space expansion of $C$ is
\begin{align}
    C_{ij}^\mu(\kp{k},\vect{0}) = &
    \sum_{\vect{R}}e^{i\kp{k}\cdot\vect{R}}
    C_{i(\vect{R})j(\vect{0})}^{\mu(\vect{0})}.
\end{align}
The real space RI expansion coefficients can be derived as
\begin{equation}
    C_{i(\vect{0})j(\vect{R})}^{\mu(\vect{0})} = \, 
    \sum_{\vect{R}'=\vect{0},\vect{R}}
    (i(\vect{0})j(\vect{R})|\nu(\vect{R}'))
    V_{\nu(\vect{R}')\mu(\vect{0})}^{-1}.
\end{equation}
Note that the translation invariance for the expansion coefficient is used in Eq. \ref{eq:co_rilvl} (i.e. $C_{i(\vect{R})j(\vect{R}')}^{\mu(\vect{R}'')} = C_{i(\vect{R}-\vect{R}'')j(\vect{R}'-\vect{R}'')}^{\mu(\vect{0})}$) .
Combing Eqs. \ref{eq:co_ri} and \ref{eq:co_rilvl} leads to
\begin{equation}
    C_{ij}^\mu(\kp{k},\kp{q}) = 
    C_{ij}^\mu(\vect{-k},\vect{0}) +
    C_{ji}^\mu(\kp{q},\vect{0}).
\end{equation}
The key point is RI-LVL decouples the $\kp{k}$,$\kp{q}$ dependence in the RI coefficients, $C_{ij}^\mu(\kp{k},\kp{q})$, and 
only a single Bloch vector index (or equivalently single lattice vector index) needs to be evaluated for the three-orbital expansion coefficients.

\subsubsection{\label{subsec:EXX-r} Exchange operator in real space for extended systems}

While the reciprocal-space representation is widely used in electronic structure methods for extended periodic systems, a drawback of using the RI-LVL approach in reciprocal space is that the expansion coefficients $C_{ij}^\mu(\kp{k},\vect{0})$ would not be sparse in our NAO-based implementation. In particular, they cannot be readily screened based on distance using the locality of the basis functions \cite{levchenko2015hybrid}. 
Alternatively, formulating the exchange operator in real space is more appealing, since the overlap between atomic basis functions pairs becomes negligible with increasing distance, and the density matrix component $D_{lm}\spin$ likewise decays with increasing distance. 
The real-space representation of the exchange operator can be related to that in the reciprocal space (see Eq. \ref{eq:exchange_nao_density}) as
\begin{equation}
    K_{ij}\spin(\kp{k},t)=\sum_{\vect{R}}
    e^{i\kp{k}\cdot\vect{R}}X_{ij}\spin(\vect{R},t),
\end{equation}
where $X_{ij}\spin(\vect{R},t)$ is the real-space representation given by
\begin{equation}
\begin{aligned}
    X_{ij}\spin(\vect{R},t) =  &
    \sum_{lm}\sum_{\vect{R}'\vect{R}''}
    D_{lm}\spin(\vect{R}',t) \iint \frac{1
    }{
    \left|\vect{r}-\vect{r}'\right|
    } \phi_i(\vect{r})
    \phi_l(\vect{r}-\vect{R}'')\\
    & \phi_m(\vect{r}'-\vect{R}'-\vect{R}'')
    \phi_j(\vect{r}'-\vect{R}) d\vect{r}\,d\vect{r}'.
\end{aligned}
\end{equation}
Because of translation symmetry, summing $\phi_i(\vect{r})$ over all unit cells would repeatedly count equivalent contributions, causing the periodic sum to diverge. Therefore, only the basis functions $\phi_i(\vect{r})$ belonging to the $\vect{0}$ unit cell are considered.
The density matrix in real-space is related to its reciprocal-space representation by
\begin{equation}
    D_{lm}\spin(\vect{R},t) =
    \frac{1}{N_{\kp{k}}}\sum_{\kp{k}}
    e^{-i\kp{k}\cdot\vect{R}}
    D_{lm}\spin(\kp{k},t).
\end{equation}
Using RI, the real space exchange matrix is expressed as
\begin{equation}
\begin{aligned}
    X_{ij}\spin(\vect{R},t) = &
    \sum_{l\vect{R}'}
    \sum_{m\vect{R}''}
    \sum_{\mu\vect{Q}'}
    \sum_{\nu\vect{Q}''}
    D_{lm}\spin(\vect{R}+\vect{R}''-\vect{R}',t)
    C_{i(\vect{0})l(\vect{R}')}^{\mu(\vect{Q}')} \\
    & V_{\mu(\vect{0})\nu(\vect{R}+\vect{Q}''-\vect{Q}')}
    C_{j(\vect{0})m(\vect{R}'')}^{\nu(\vect{Q}'')}
\end{aligned}
\label{eq:X_R}
\end{equation}
where $\vect{Q}'$ and $\vect{Q}''$ are also lattice vectors. 
For the case of RI-LVL (Eq. \ref{eq:ap_rilvl}), the choice of $\vect{Q}'$ is either $\vect{0}$ or $\vect{R}'$, and the choice of $\vect{Q}''$ is either $\vect{0}$ or $\vect{R}''$.
The auxiliary basis functions $\mu$ and $\nu$ are restricted to those centered on the atom of either basis $\phi_i,\phi_l$ and $\phi_j,\phi_m$, respectively.
With the aforementioned advantages offered through the real-space formula,
the sum over lattice vectors $\vect{R}'$ and $\vect{R}''$ is limited by the locality of the NAO basis overlaps.
The choice of $\vect{R}$ is given by the decay of density matrix or the slow $1/\vect{R}$ decay of the the Coulomb matrix.
\citeauthor{goedecker1999linear} has shown the density matrix decays exponentially in insulators but only polynomially in metals\cite{goedecker1999linear}. 
As a result, the real space formula, which works great for evaluating RI coefficients $C_{ik}^\mu$, is not ideal for the Coulomb matrix $V_{\mu\nu}$.
If one instead considers Born–von Karman (BvK) periodic boundary conditions, the Coulomb matrix can be obtained by the inverse Fourier transformation of its reciprocal space representation
\begin{equation}
    \label{eq:Coulomb_BvK}
    V_{\mu\nu(\vect{R})} = 
    \frac{1}{N_{\kp{k}}}\sum_{\kp{k}}
    e^{-i\kp{k}\cdot\vect{R}}
    \tilde V_{\mu\nu}(\kp{k}),
\end{equation}
where the lattice vectors $\vect{R}$ are restricted to the BvK cell, and the explicit origin unit cell $\vect{0}$ is omitted for brevity. 
The Coulomb matrix can extend beyond the BvK PBC by considering the periodicity of the Coulomb matrix
\begin{equation}
V_{\mu\nu(\vect{R})} = V_{\mu\nu(\vect{R}+\vect{R}^{\text{BvK}})},
\end{equation}
where $\vect{R}^{\text{BvK}}$ is a lattice vector of the BvK superlattice. 
An efficient way of evaluating Eq.~\ref{eq:Coulomb_BvK} is using the cut-Coulomb operator, defined as
\begin{equation}
    \label{eq:Vcut}
    V_\text{cut}(\vect{r}) = 
    \frac{1}{2\left|\vect{r}\right|}
    \text{erfc}\left[
    \eta\left(\left|\vect{r}\right|-r_\text{cut}\right)
    \right],
\end{equation}
where $r_\text{cut}$ is the decay radius, and $\eta$ is the inverse decay width \cite{levchenko2015hybrid, spencer2008efficient}.
These parameters are set in FHI-aims code with a default value of $8~\text{\AA}$ for $r_\text{cut}$, and $\eta$ determined by the integration grid \cite{blum2009ab, abbott2025roadmap}. 
The general reciprocal space cut-Coulomb operator is given by
\begin{equation}
\begin{aligned}
    \tilde V_{\mu\nu}&(\kp{k}) = \\
    & \sum_{\vect{R}}
    e^{i\kp{k}\cdot\vect{R}}
    \iint P_\mu(\vect{r})
    V_\text{cut}(\left|\vect{r}-\vect{r}'\right|)
    P_\nu(\vect{r'}-\vect{R})d\vect{r}\,d\vect{r}'.
\end{aligned}
\end{equation}
where $P_\mu$ and $P_\nu$ are auxiliary basis functions, with $P_\mu$ centered in the origin unit cell and $P_\nu$ centered in the cell displaced by $\vect{R}$. 
It is worth mentioning that the cut-Coulomb operator provides a natural representation for the Coulomb interaction of range-separated hybrid functionals, for which the electron–electron interaction is partitioned according to its range.

\subsubsection{\label{subsec:EXX-practice} Implementation Details in FHI-aims}
Because the contraction in Eq.~\ref{eq:X_R} involves large, irregularly shaped matrices (and, in practice, high–dimensional tensor blocks), careful data layout and parallel work distribution are essential to prevent memory pressure and communication overhead.
In practice, the RI coefficients are grouped according to the atom pairs of the auxiliary basis.\cite{bussy2024efficient,kokott2024efficient}
The exchange matrix associated with atoms $A_1^1$ and $A_1^2$, where the superscript denotes the pair and the subscript denotes the atom within that pair, is expressed as 
\begin{equation}
\begin{aligned}
    & X_{i\in A_1^1,j\in A_1^2}\spin(\vect{R}) = 
    \sum_{A_2^1(\vect{R}')}
    \sum_{\small \substack{k\in A_2^1(\vect{R}') 
    \\ \nu\in A_1^2(\vect{R})}}
    F_{ik}^{\nu(\vect{R})}
    E_{jk(\vect{R}-\vect{R}')}\spin^{\nu} \\
    & +
    \sum_{A_2^2(\vect{R}'')}
    \sum_{\small \substack{l\in A_2^2(\vect{R}'') 
    \\ \nu\in A_2^2(\vect{R}'')}}
    \left[
    2G_{il}\spin^{\nu}\left(\vect{R}+\vect{R}''\right) +
    H_{il}\spin^{\nu}\left(\vect{R}+\vect{R}''\right)
    \right]
    C_{lj(-\vect{R}'')}^{\nu},
    \label{eq:X_implement}
\end{aligned}
\end{equation}
where
\begin{align}
    F_{ik(\vect{R}')}^{\nu(\vect{R})} = & 
    \sum_{\mu\in A(i)}
    C_{ik(\vect{R}')}^\mu
    V_{\mu\nu(\vect{R})}, \label{eq:F_mat}\\
    E_{jk(\vect{R})}\spin^{\nu} = & 
    \sum_{l\vect{R}''}
    C_{jl(\vect{R}'')}^\nu
    D_{kl}\spin(\vect{R}+\vect{R}''), \label{eq:E_mat} \\
    G_{il}\spin^{\nu}(\vect{R}) = & 
    \sum_{\mu\in A(i)}
    \left(E_{il(-\vect{R})}\spin^{\nu}\right)^*
    V_{\mu\nu(\vect{R})}, \label{eq:G_mat}\\
    H_{il}\spin^{\nu}(\vect{R}) = & 
    \sum_{k\vect{R}'}
    F_{ki(\vect{-R}')}^{\nu(\vect{R}-\vect{R}')}
    D_{kl}\spin(\vect{R}-\vect{R}') \label{eq:H_mat}. 
\end{align}
Here, $A(i)$ in Eqs. \ref{eq:F_mat} and \ref{eq:G_mat} is the atom where basis function $i$ is centered, 
$\mu \in A(i)$ indicates an auxiliary basis function $P_\mu(\vect{r})$ on atom $A(i)$, 
and $^*$ in Eq. \ref{eq:G_mat} represents the complex conjugate. 
Considering the translational symmetry of the RI coefficient, the first $F$ and $E$ terms summing over $A_2^1$, 
the second part (including $G$)
and the last part with $H$ recover the $\kp{Q}'=\kp{Q}''=\vect{0}$ , 
$\kp{Q}'= \vect{R}',\kp{Q}''=\vect{0};
\kp{Q}'= \vect{0},\kp{Q}''=\vect{R}''$, 
and $\kp{Q}'= \vect{R}',\kp{Q}''=\vect{R}''$ component in Eq. \ref{eq:X_R}, respectively.

The use of the intermediate matrices ($E,F,G$ and $H$) is to minimize the number of matrix multiplications.
The matrices $E$ and $F$ are sparse.
The indices $i$ and $k(\vect{R}')$ of matrix $F$ are restricted by the overlap of corresponding basis functions.
When there is no overlap between $i$ and $k(\vect{R}')$, the multipole moments of $\nu(\vect{R})$ are used to determine the matrix elements of $F$. 
A unitary transform is then performed on the auxiliary basis functions of each atom and in each $lm$ angular momentum channel such that the multipole moments vanish for all but one of the auxiliary basis functions\cite{Betzinger_2010_FLapw}. 
In matrix $E$ the indices $j$ and $k(\vect{R}')$ are bounded by the decay of density matrix, which is system dependent.
Additionally, a screening parameter $\epsilon$ can be used to screen all the small matrix elements used in Eq. \ref{eq:X_implement}.
 
The hybrid XC functional contribution to the real-time Kohn–Sham Hamiltonian retains the same formal structure as the hybrid XC contribution in ground-state DFT \cite{levchenko2015hybrid,kokott2024efficient}. Other implementation details, including the message passing interface (MPI) parallelization and memory distribution, are discussed in Ref.\citenum{levchenko2015hybrid} and \citenum{kokott2024efficient}.
The main distinction for the RT-TDDFT context is primarily numerical rather than theoretical: the time-dependent orbitals and density matrix are complex-valued, and the nonlocal exchange operator must be reconstructed at each propagation step.
These aspects require complex arithmetic, explicit enforcement of Hermiticity, and efficient parallel distribution of the exchange build.
The implementation discussed above extends the existing RI-LVL hybrid framework in \texttt{FHI-aims} code to support complex data and dynamic updates during time evolution.
The key difference is that, in RT-TDDFT, we exploit Hermiticity by evaluating the symmetric (real) and antisymmetric (imaginary) components separately, thereby reusing and optimizing algorithms originally developed for real-valued hybrid-XC DFT.
As a result, this RT-TDDFT implementation readily takes advantage of recent\cite{kokott2025efficient} and future improvement on hybrid XC functional implementations for ground-state DFT.
While we do not repeat the scaling tests for hybrid XC functionals as detailed in Ref. \cite{kokott2024efficient}, {we provide the timing data for individual application cases in the following section}.


\section{\label{sec:app} Validation and demonstration}

\subsection{\label{subsec:water} Isolated Water molecule}

\begin{figure*}
    \centering
    \includegraphics[width=0.95\linewidth]{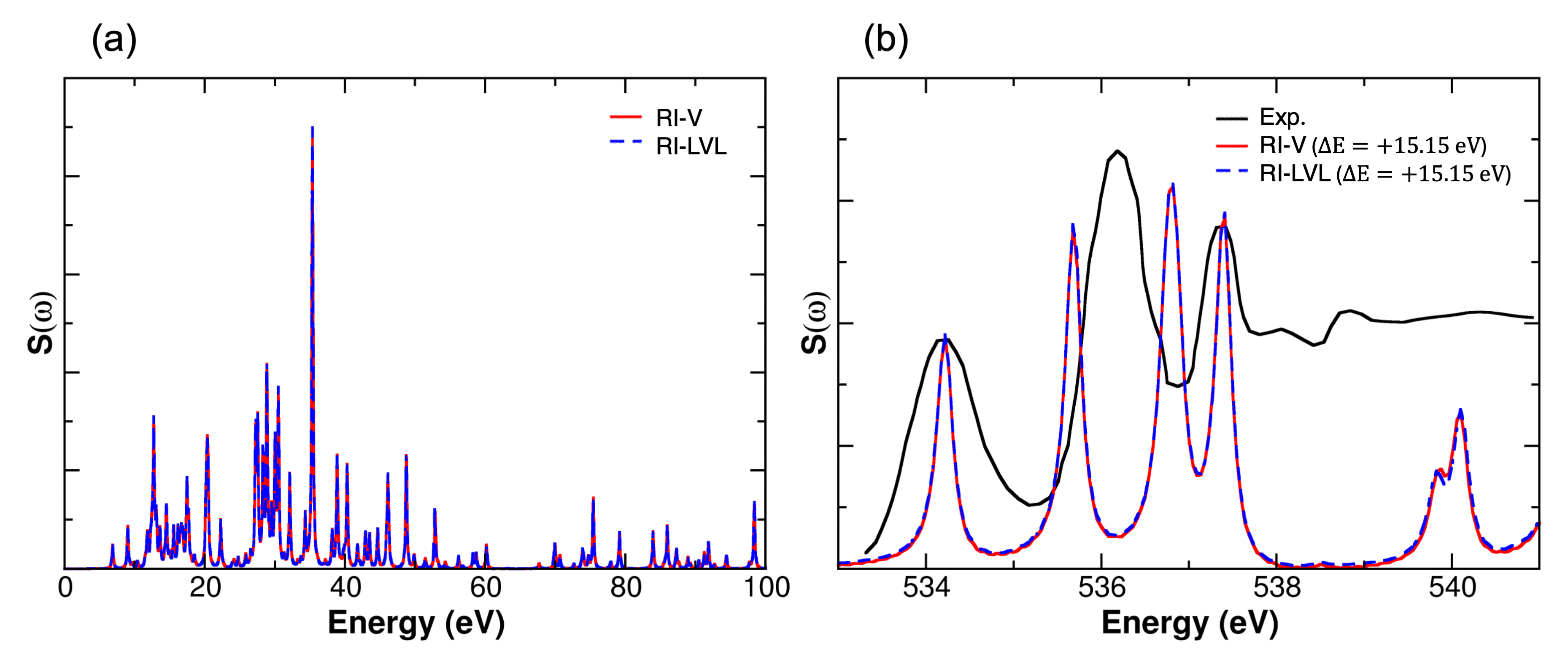}
    \caption{(a) Valence-electron dipole strength function for a water molecule computed using both the RI-V and RI-LVL approaches with the B3LYP XC functional\cite{Stephens_1994_B3LYP}. RI-LVL is shown with a dashed line to ensure visibility of the RI-V curve. (b) $K$-edge core absorption spectra for the RI-V and RI-LVL approaches, compared with experiment (Exp) from Ref. \citenum{Wilson_2001_Expwaterspec}. The calculated spectra are shifted such that the lowest $K$-edge peak matches that of experiment.}
    \label{fig:water_Vcomp}
\end{figure*}

To validate our implementation, we first compare the optical absorption spectrum for an isolated water molecule obtained using the RI-LVL framework to that obtained using the conventional RI-V framework.
Using the B3LYP hybrid functional\cite{Stephens_1994_B3LYP}, we employed the length gauge (Eq.~\ref{eq:length_gauge}) and applied a weak $\delta$-kick at $t=0$, in the form of a constant electric field of magnitude 5.1 $\times$ $10^8$ V/m, to excite a broad range of frequencies.
The field was subsequently turned off, and the RT-TDDFT states were propagated for a total simulation time of 30.0 fs using a time step of 0.00012 fs (0.005 a.u.) and the Crank-Nicholson propagator. \cite{Crank_Nicolson_1947, marques2004time}
The \texttt{tight+aug2} basis was used, as it has been shown to provide good convergence with respect to larger benchmark basis sets for molecules \cite{hekele2021all, Liu2020atomsbse}
Specifically, it combines a dense numerical integration grid with $s$- and $p$-type Dunning augmentation functions. 
Through \textsc{FHI-aims} all-electron framework, we are able to compare the valence and core excitations on equal footing using the same computational procedure.
To ensure accurate calculation of the core spectra, relativistic corrections were included in the TD-KS Hamiltonian through the specific atomic zero-order regular approximation (atomic ZORA) approach\cite{blum2009ab}, while the small integration time step was chosen so that the associated high-energy excitations could be resolved.
This procedure was repeated for each Cartesian direction to obtain the time-dependent electronic dipole moment 
\begin{equation}
\begin{aligned}
    \boldsymbol{\mu}(t)= \int \vect{r} \,\rho(\vect{r},t)\,d\vect{r}. 
    \label{eq:dipole}
\end{aligned}
\end{equation}
Since the time-dependent current is defined as the change in the electronic dipole with respect to time 
\begin{equation}
\begin{aligned}
    \mathbf{J}(t)=\frac{d\boldsymbol{\mu}(t)}{dt},
    \label{eq:polarization}
\end{aligned}
\end{equation}
the electronic dipole can be used to obtain the polarizability tensor $\alpha_{ij}(\omega)$ 
\begin{equation}
\begin{aligned}
    \alpha_{ij}(\omega)= \frac{\int_0^{\infty} e^{i\omega t}{\mu}_i(t)e^{-t/\tau}dt}
    {\!\int_0^{\infty} e^{i\omega t}
  E_j(t)\,dt},
    \label{eq:polarizability}
\end{aligned}
\end{equation}
where $i$ refers to the direction of the time-dependent dipole, $j$ refers to the direction of the $\delta$-kick, and $\tau$ is a damping coefficient or function which suppresses truncation artifacts arising from the finite simulation time and simultaneously broadens the spectral peaks.
The polarizability tensor is in turn used to calculate the dipole strength function $S(\omega)$ \cite{Yabana_1999_dipolestength}, 
\begin{equation}
\begin{aligned}
    S(\omega)= \frac{2\, \omega}{3\, \pi} \, \mathrm{Im}\{\mathrm{Tr[\alpha_{ij}(\omega)]\}}.
    \label{eq:dipole strenght function}
\end{aligned}
\end{equation}
In the case of isolated systems, the dipole strength function gives the optical absorption as the macroscopic dielectric function is not well defined. 

Figure \ref{fig:water_Vcomp}(a) shows the resulting dipole strength function, computed with a damping constant of 6~fs, as a function of energy for the valence excitation of a water molecule. 
The RI-LVL approach gives an exact match to the RI-V result over the entire energy range shown.
Convergence tests, with respect to time step, damping coefficient and total simulation time, are included in the Supporting Information (Section I.A).
Additionally, we find the agreement between the two RI approaches to be independent of simulation time, damping constant, and XC functional (see Supporting Information, Sections I.B.1–I.B.3). 
Exact agreement was also observed for the standard \texttt{tight} and \texttt{intermediate} basis set settings of \textsc{FHI-aims} \cite{abbott2025roadmap, blum2009ab}.
For the \texttt{light} basis, full agreement was observed when a single auxiliary basis function was included for hydrogen for the RI-LVL density fitting (see Supporting Information Section I.B.4). 

Figure \ref{fig:water_Vcomp}(b) shows that an exact match between the two RI approaches was also observed for the $K$-edge (excitation from oxygen 1$s$ electrons) region of the spectra. 
Both calculated spectra have been blue shifted by 15.15 eV (i.e. shifted such that the lowest peak matches the experimental analog\cite{Wilson_2001_Expwaterspec}). 
We note that the shape and magnitude of the \textit{K}-edge spectra peaks are independent of the time step, whereas the applied shift does depend on the time step used (see Supporting Information Section I.A.).
The relativistic correction was found to have no effect on the valence spectra but shifted the \textit{K}-edge excitation energies by $+0.8~\mathrm{eV}$; the same for both RI approaches and in line with previous work \cite{Keller_2020_RelavisticGW,hekele2021all}.
Additionally, while the use of the PBE semi-local XC functional was found to give a similar $K$-edge spectrum, it required a larger shift of 24.1 eV to match the lowest-energy experimental peak (see Supporting Information Section 1.C.1).

\begin{figure}
    \centering
    \includegraphics[width=0.95\linewidth]{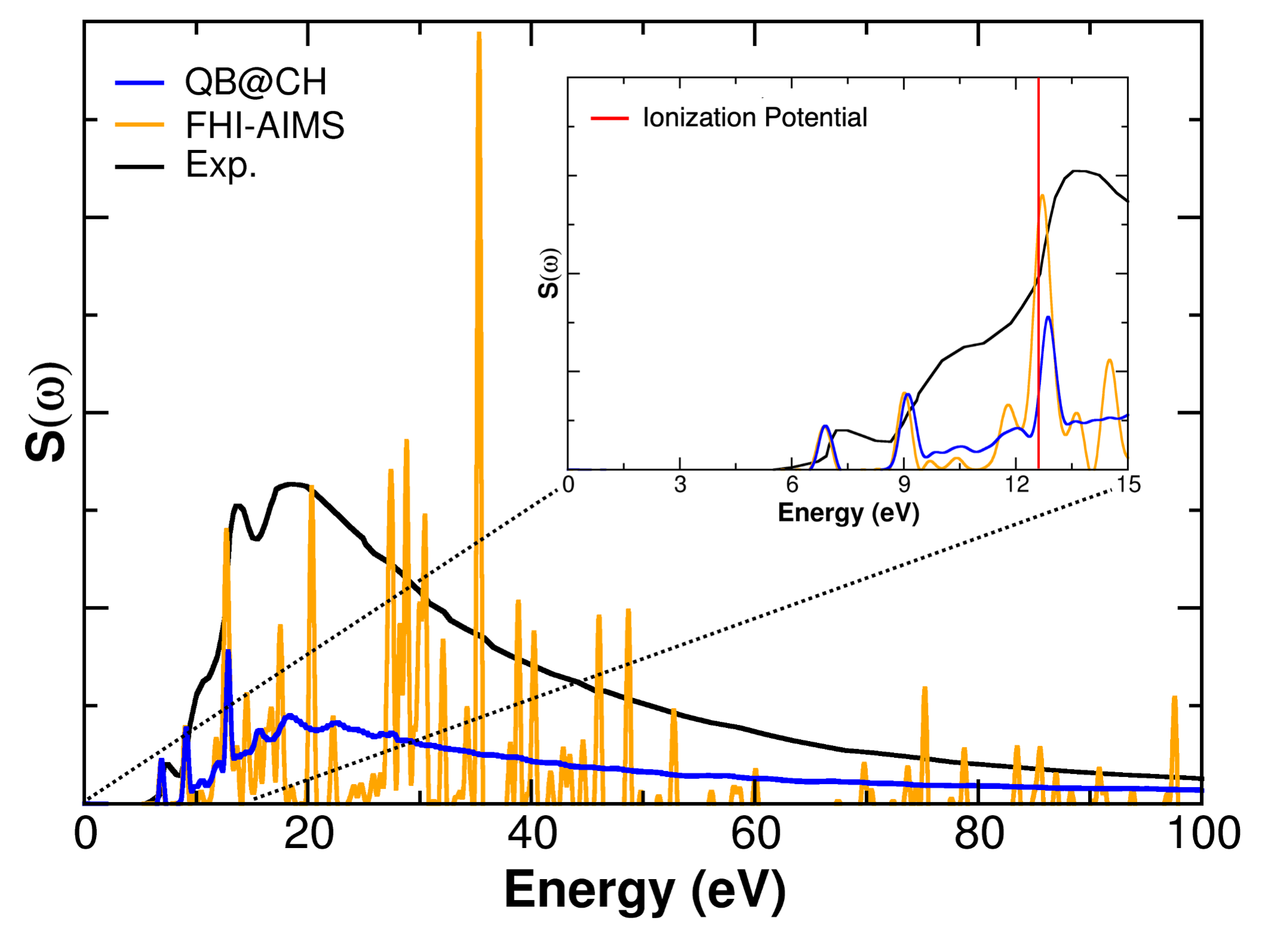}
    \caption{Comparison of the valence-electron dipole strength function obtained using the NAO basis in \textsc{FHI-aims} and a plane-wave basis in QB@CH \cite{schleife2012plane, Shepard_2024_EXX}, together with experiment (Exp)\cite{Hayashi_2000_IPwater}. Both theoretical spectra were calculated using the B3LYP XC functional. The experimental spectrum was scaled such that its lowest-energy peak has the same intensity as the corresponding peak in the two theoretical spectra. While the spectra differ at higher excitation energies, excellent agreement is observed between the two theoretical spectra in the bound-state region below the ionization potential\cite{Hayashi_2000_IPwater} (red line), as highlighted in the inset.}
    \label{fig:water_compPW}
\end{figure}

Having established that the RI-LVL implementation reproduces the RI-V approach exactly for both valence and core excitations, we next assess the limitations imposed by the underlying NAO basis itself. 
Although an NAO basis allows valence and core excitations to be treated on an equal footing, unlike plane-wave pseudopotential (PW-PP) methods which do not explicitly treat core electrons, localized atom-centered basis sets face a complementary challenge: for electronic excitations involving ionization, they are not as well suited to represent the resulting diffuse excitations.
We note that this limitation is actively being addressed through approaches such as atomic Gaussian basis sets optimized for the continuum.\cite{Leonetti_26_ionize}
In this context, comparison with plane-wave calculations is particularly useful, as plane-wave basis sets provide a more natural description of ionization processes.
To this end, we compare the single-water spectra obtained from \textsc{FHI-aims} and the PW-PP code QB@CH\cite{Shepard_2024_EXX}\yk{reminder to cite 2014 CiSE in revision.}, together with experiment.
For the \textsc{FHI-aims} calculation, the RI-LVL approach was used, and all other parameters were kept the same as the calculation described above, except for a shorter run time of 9.7 fs and a larger time step of 0.0024 fs (0.1 a.u.). 
The QB@CH spectrum was calculated using a 60 Rydberg cutoff for PW expansion, ONCV pseudopotentials (PP) \cite{Hamann2013ONCV, Gygi2015ONCV} and a 30 Bohr cell with a complex absorbing potential to prevent interactions with periodic images. 
A delta-kick was applied using the length gauge, identical in magnitude to that used in the \textsc{FHI-aims} simulation, and the system was propagated using the same 9.7~fs run time and 0.0024~fs time step as the \textsc{FHI-aims} calculation, along with the ETRS propagator. 
Since the use of valence-only pseudopotentials excludes core-electron excitations, the valence excitations can be fully resolved with this larger time step and shorter run time.
A damping constant of 5~fs was used for the broadening of both spectra. 

Figure \ref{fig:water_compPW} shows the resulting single-water spectra obtained from the \textsc{FHI-aims} and QB@CH codes, alongside experiment \cite{Hayashi_2000_IPwater}.
The experimental spectrum was scaled such that the lowest excitation peak (at $\sim$7~eV) has the same magnitude as the two computed spectra. 
Over the broad energy range from 0 to 100 eV, the dipole strength function calculated using \textsc{FHI-aims} does a poor job of reproducing the overall spectral features while QB@CH captures the general features of the experimental spectrum well despite noticeable differences in the peak intensities beyond the first excitation.
However, when we focus on the region below the ionization potential (indicated by the red line in the inset at 12.6 eV \cite{Hayashi_2000_IPwater}), the \textsc{FHI-aims} spectrum performs comparably to the PW-PP calculation in reproducing the peak positions and relative intensities. 
In addition, although the peak positions do not exactly match experiment, the use of the B3LYP XC functional improves the agreement by approximately 0.5 eV relative to the PBE spectra (see Supporting information Section 1.C.2), highlighting the importance of hybrid XC functionals. 
Overall, these results for isolated water validate our RI-LVL implementation, as exact agreement with the RI-V approach was observed for both valence and core excitations.
Comparison with PW-PP calculations further shows the expected limitations of the localized atom-centered orbital basis for ionizing excitations, while bound-state excitations are accurately described.

\begin{figure*}
    \centering
    \includegraphics[width=0.99\linewidth]{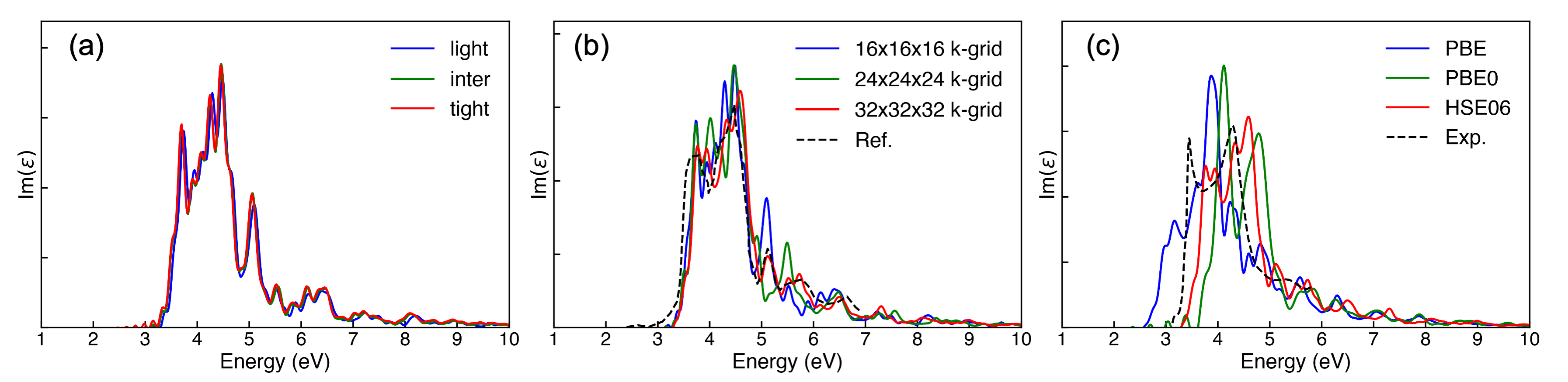}
    \caption{The imaginary parts of the dielectric functions of Si. (a) The \texttt{light}, \texttt{intermediate}, and \texttt{tight} NAO basis using a $16\times16\times16$ $\kp{k}$-point mesh with the HSE06 functional. 
    (b) The $\Gamma$-centered $16\times16\times16$, $24\times24\times24$, and $32\times32\times32$ $\kp{k}$-point meshes using the \texttt{light} NAO basis with the HSE06 functional, compared with $30\times30\times30$ $\kp{k}$-point mesh HSE06 results (Ref) from Ref. \citenum{pemmaraju2019valence}.
    (c) The PBE, PBE0, and HSE06 functionals, using the \texttt{light} NAO basis and a $32\times32\times32$ $\kp{k}$-point mesh, compared with experimental data (Exp) from Ref. \citenum{lautenschlager1987temperature}.
    }
    \label{fig:si}
\end{figure*}

\subsection{Extended systems}
Having validated the RI-LVL implementation with isolated water, we extend our analysis to the main goal of RI-LVL, extended systems. 
The macroscopic current $\vect{J}(t)$ yields the dielectric function and other optoelectronic response properties, and it is given by
\begin{align}
    \vect{J}(t)&  = -\frac{1}{V}\int  \vect{j}(\vect{r},t)\,d\vect{r} 
\end{align}
where $V$ is the simulation cell volume and $\vect{j}(\vect{r},t)$ is the induced current density. 
The current density can be obtained from the kinetic momentum operator as
\begin{align}
    \vect{j}(\vect{r},t) & = \frac{1}{2}\sum_{n}\int_{\mathrm{BZ}} \,d\kp{k}\, 
  f\spin_{n\kp{k}}
  \left[\psi\spin_{n\kp{k}}^*(\vect{r},t)(-i\vect{\nabla}+\vect{A}(t))\psi\spin_{n\kp{k}}(\vect{r},t) + c.c. \right].
\end{align}
Its Fourier transform yields the frequency-dependent conductivity tensor,
\begin{equation}
  \sigma_{ij}(\omega) = \frac{
  \!\int_0^{\infty} e^{i\omega t}
  {J_i}(t)e^{-t/\tau}\,dt}
  {\!\int_0^{\infty} e^{i\omega t}
  E_j(t)\,dt}
\end{equation}
where $\tau$ is a phenomenological damping constant mimicking finite lifetimes and experimental broadening, $i$ refers to the direction of the time-dependent current, and $j$
refers to the direction of the perturbing field.
The dielectric function is given by
\begin{equation}
  \epsilon(\omega) = 1+i\frac{4\pi}{3\omega}\text{Tr}[\mathbf{\sigma}(\omega)].
  \label{eq:dielectric_conductivity}
\end{equation}

\subsubsection{\label{subsec:Si} Crystalline silicon}

We use bulk crystalline silicon as a benchmark system for the hybrid-functional RT-TDDFT implementation.\cite{pemmaraju2019valence}
The two-atom unit cell is modeled using a lattice parameter of 5.402~\AA.
We investigate the convergence of the calculated dielectric function with respect to the NAO basis set, Brillouin zone sampling, and XC functional.
In particular, the convergence behavior of \texttt{light}, \texttt{intermediate}, and \texttt{tight} NAO basis set settings\cite{blum2009ab} are reported.
The Brillouin-zone integration is performed using $\Gamma$-centered $16\times16\times16$, $24\times24\times24$, and $32\times32\times32$ $\kp{k}$-point meshes here without using any symmetry reduction schemes for generality.
For the XC approximations, the PBE\cite{perdew1996generalized}, PBE0\cite{adamo1999toward}, and HSE06\cite{heyd2003hybrid} XC functionals are employed.
A constant vector potential corresponding to a $\delta$-kick perturbation was applied starting from $t=0$ with a strength of $5.1\times10^8$ V/m. 
For the time propagation, we employed the exponential midpoint propagator with a time step of 0.0048~fs (0.2 a.u. time) for a total simulation time of 24.2~fs. 
Additional computational details and convergence tests, including the dependence on the simulation time and damping constant, are provided in the Supporting Information. (see Section II).

\par Using 384 MPI processes evenly distributed across 4 AMD (EPYC 9654/EPYC 9684X) nodes of the Sycamore Cluster at the University of North Carolina at Chapel Hill (UNC), the $16\times16\times16$ $\kp{k}$-point mesh system takes 0.09~s per time step for PBE and 8.6~s per time step for HSE06 XC functionals. 
While computational performance such as the runtime tends to depend strongly on the physical architectures, including system interconnect/fabric in modern high-performance computers, the two orders-of-magnitude increase in the computational cost is typical for these condensed matter systems, as also seen for the RT-TDDFT based on PW-PP formalism\cite{Shepard_2024_EXX}. 
The implementation of efficient hybrid XC functionals has been already discussed in the context of ground-state DFT in Ref. \citenum{kokott2024efficient}, and we refer the reader there for detailed timing and performance discussion.

\par Figure~\ref{fig:si} shows the imaginary part of the frequency-dependent dielectric function, corresponding to the optical absorption spectrum, obtained from the time-dependent electronic current density using Eq.~\ref{eq:dielectric_conductivity} with a damping constant of 20~fs.
For Si atoms, the default 2020 settings in \textsc{FHI-aims} code employ the \texttt{tier 1} NAO basis set for both the \texttt{light} and \texttt{intermediate} numerical settings. 
This basis consists of the minimal free-atom basis augmented by one additional \(s\), \(p\), \(d\), and \(f\) function each. 
The \texttt{intermediate} settings differ from the \texttt{light} settings primarily through the use of a denser numerical integration grid. 
The \texttt{tight} settings use the same integration grid as the \texttt{intermediate} settings, but further augment the basis with one additional \(d\) and \(g\) function each.
As shown in Figure \ref{fig:si}(a), 
no significant dependence on either the basis set or numerical integration settings is observed. 
These results indicate that the default 2020 \texttt{light} settings in the \textsc{FHI-aims} code are already well converged for RT-TDDFT calculation of the dielectric function of crystalline Si with the HSE06 functional.

Figure~\ref{fig:si}(b) compares the convergence with respect to the $\Gamma$-centered $\mathbf{k}$-point meshes using the \texttt{light} NAO basis and the HSE06 functional.
All spectra exhibit similar features, including major peaks near 3.7 and 4.4~eV, as well as weaker features around 5.1, 5.5, and 6.5~eV.
These results show a good agreement with the reference calculation by \citeauthor{pemmaraju2019valence}, which employed a $30\times30\times30$ $\mathbf{k}$-point mesh.\cite{pemmaraju2019valence}

The comparison between different XC functionals, using the \texttt{light} NAO basis and a $32\times32\times32$ $\mathbf{k}$-point mesh, is shown in Fig.~\ref{fig:si}(c).
The PBE results exhibit a main peak at approximately 3.8~eV and a shoulder near 3.0~eV.
The additional peaks near 4.3 and 5.6~eV are artificial, arising from insufficient $\mathbf{k}$-point sampling; they disappear with denser $\mathbf{k}$-point meshes, as shown in the Supporting Information.
Compared to PBE, the PBE0 results show an overall blue shift, with a dominant peak near 4.1~eV and weaker peaks around 3.8 and 4.7~eV.
Unlike the PBE case, these PBE0 spectral features remain largely unchanged upon increasing the $\mathbf{k}$-point sampling from $16\times16\times16$ to $32\times32\times32$.
As discussed above, the HSE06 results show major peaks near 3.7 and 4.4~eV, and the overall spectral shape agrees well with experimental measurements, except for an approximately 0.2~eV blue shift.
These comparisons highlight the importance of hybrid XC functionals, particularly range-separated hybrid XC functionals, for accurately describing the optical spectrum of bulk Si\cite{refaely2015solid,wing2019comparing}.

\subsubsection{\label{subsec:SiC} Crystalline silicon carbide}

\begin{figure*}
    \centering
    \includegraphics[width=0.99\linewidth]{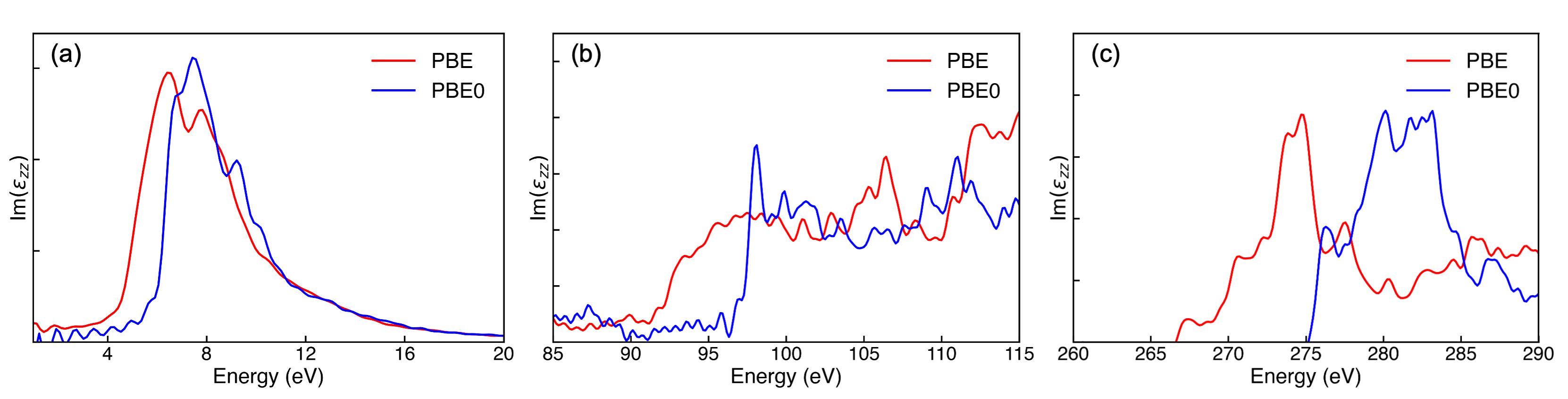}
    \caption{The \textit{zz}-component of the imaginary dielectric function for 2H-SiC calculated using the PBE and PBE0 XC functionals and the \texttt{light} basis setting in \textsc{FHI-aims}. The (a) valence (b) \textit{L}-edge of silicon and (c) \textit{K}-edge of carbon energy regions are plotted.}
    \label{fig:SiC}
\end{figure*}

In addition to the optical/UV-vis absorption spectra due to valence electronic excitation, 
the all-electron implementation based on the NAO basis allows us to compute X-ray absorption (XAS) spectra of extended periodic systems. 
To demonstrate this capability we calculate the valence and XAS spectra for the 2H-SiC polytype of crystalline silicon carbide. 
The 2H polytype forms a hexagonal unit cell with four atoms, and we used experimentally determined lattice parameters of \textit{a}~=~3.076~\AA \ and \textit{c}~=~5.048~\AA \cite{PEMMARAJU_2019_velocitygauge}. 
A \textbf{k}-point mesh of $16\times16\times12$ was found to be sufficient for Brillouin-zone integration (see Supporting Information Section III.A). 
For the basis set we used the \texttt{light} setting within \textsc{FHI-aims}, as previous results showed this basis set setting gave converged results for the imaginary part of the dielectric function\cite{hekele2021all}. 
The \texttt{light} setting for both atoms consists of the minimal free-atom basis plus one additional \textit{s, p, d} and \textit{f} function for silicon and one additional \textit{s, p} and \textit{d} function for carbon. 

For the calculation of the spectra, the velocity-gauge RT-TDDFT formalism was used. A constant vector potential corresponding to an electric field of strength $5.1\times10^8$ V/m was applied starting at $t=0$. The semi-local PBE \cite{perdew1996generalized} and hybrid PBE0 \cite{adamo1999toward} XC functionals were compared here. To accurately describe the core spectra, relativistic effects were included using the atomic ZORA approach \cite{blum2009ab}. The electronic wavefunctions were propagated using the Crank--Nicholson propagator with a time step of 0.00024 fs (0.01~a.u.)\cite{PEMMARAJU_2019_velocitygauge}, and the simulations were run for 6.6 fs. The system was perturbed only along the z-direction to obtain the \textit{zz}-component of the dielectric function. The time-dependent electronic current was subsequently used to calculate the absorption spectrum using Eq. \ref{eq:dielectric_conductivity}, with a damping constant of 2.5~fs. 
Using 192 MPI processes on one AMD node ((EPYC 9654/EPYC 9684X) of the UNC Sycamore Cluster, the $16\times16\times12$ $\kp{k}$-point mesh system takes 1.1~s per time step for the PBE and 174.5~s per time step for the PBE0 XC functionals.
Additionally, convergence tests, with respect to simulation time and damping coefficient, are provided in the Supporting Information (see Section III.B).

Figure \ref{fig:SiC} shows the calculated imaginary dielectric function of silicon carbide, divided into three energy windows: (a) 1--20 eV, corresponding to valence electron excitations; (b) 85--115 eV, corresponding to the silicon \textit{L}-edge excitation; and (c) 260--290 eV, corresponding to the carbon \textit{K}-edge excitation. In the valence region, the first peak has nearly the same intensity for both XC functionals, while its position exhibits a blue shift from 6.4 eV (PBE) to 7.4 eV (PBE0). The second peak undergoes a slightly larger blue shift of 1.4 eV from PBE to PBE0, accompanied by a decrease in intensity. 

For the silicon \textit{L}-edge excitation region, shown in Figure \ref{fig:SiC} (b), the onset of the main plateau-like spectral feature exhibits a blue shift of approximately 4.5 eV, from 92.0 eV for PBE to 96.5 eV for PBE0. 
Apart from the blue shift, the general intensity of the plateau region is similar between the two XC functionals, although the leading edge of the intensity is greater when using PBE0. 
For the carbon \textit{K}-edge optical excitation region, a larger blue shift of 8.5 eV is observed for the PBE0 spectrum. 
Aside from this blue shift, both XC functionals produce similar spectral features.
While the ZORA correction produces no discernible changes to the valence or silicon \textit{L}-edge regions, the relativistic correction results in a 0.25 eV blue shift in the carbon \textit{K}-edge region for both XC functionals.
The overall excitation intensity is again similar between the two XC functionals, although the initial onset is greater for PBE0. 
While experimental data for 2H-SiC remains limited to make direct comparison, the calculated optical spectra provides insight into its electronic response. 
More broadly, these results demonstrate the utility of hybrid XC functionals in RT-TDDFT for studying the electronic excitation responses of periodic extended systems, including those originating from core-electron excitation.

\section{\label{sec:con} Conclusion}
In this work we presented an extension of RT-TDDFT method with the exact (Hartree-Fock) exchange for extended periodic systems within the all-electron \texttt{FHI-aims} code. 
The importance of resolution-of-identity (RI) for the four-electron integrals is discussed, along with the need for its localized version (RI-LVL) when extended systems are studied. 
We detailed the extension of the RI-LVL method to handle complex, time-dependent quantities and the dynamical updating of the exchange operator. 
The implementation is validated against the RI-V method for an isolated water molecule, demonstrating excellent agreement for both valence and core excitations.
We then demonstrate our implementation on bulk silicon, highlighting the importance of hybrid XC functionals and reproducing reference data from literature. 
Finally, we demonstrated the capability of our implementation to calculate valence and core spectra of extended systems using 2H silicon carbide. 
While the RT-TDDFT implementation is further optimized for emerging HPC architectures,
this RI-LVL method for EXX opens the door for RT-TDDFT studies of complex extended systems with state-of-the-art hybrid XC functionals using the \texttt{FHI-aims} code.
Ehrenfest dynamics based on RT-TDDFT would be a natural extension of this work. 
Unlike for the PW-PP implementation, the explicit dependence of NAO basis functions on atom positions requires further considerations, and it will be discussed in a future work.

\section*{\label{sec:ack}Acknowledgment}
We thank the Research Computing at the University of North Carolina at Chapel Hill for providing computational resources and support on the Sycamore cluster.
We acknowledge financial support from Molecular Simulations from First Principles e.V. (MS1P). 

\section*{\label{sec:Ref}References}

\bibliography{main}

\end{document}